\documentclass[sigconf]{acmart}

\usepackage{msc}
\usepackage{algorithmic}
\usepackage{graphicx}
\usepackage{acronym, comment, subcaption}
\usepackage{tabularx, pifont}
\usepackage[ruled,vlined,linesnumbered]{algorithm2e}
\usepackage{hyperref} 
\usepackage{mathtools}
\usepackage[colorinlistoftodos,textsize=small]{todonotes}
\usepackage{pdflscape}
\usepackage{lipsum}
\usepackage{multirow}
\usepackage{booktabs}

\usepackage{pgfplotstable}
\usepackage{colortbl}
\usepackage{xcolor}
\usepackage{etoolbox}

\usepackage{tikz}
\usetikzlibrary{chains,shapes.multipart}
\usetikzlibrary{shapes,calc}
\usetikzlibrary{automata,positioning}

\graphicspath{{Figures/}}
\acrodef{WLAN}{Wireless Local Area Network}
\acrodef{UN}{United Nations}
\acrodef{SDG}{Sustainable Development Goal}

\acrodef{ML}{Machine Learning}
\acrodef{SVM}{Support Vector Machine}
\acrodef{DT}{Decision Tree}
\acrodef{RF}{Random Forest}
\acrodef{BSM}{Basic Safety Message}
\acrodef{DL}{Deep Learning}
\acrodef{RNN}{Recurrent Neural Network}
\acrodef{LSTM}{Long Short-Term Memory}
\acrodef{CNN}{Convolutional Neural Network}
\acrodef{DDoS}{Distributed Denial-of-Service}
\acrodef{V2X}{Vehicle-to-Everything}
\acrodef{k-NN}{k-Nearest Neighbor}
\acrodef{LLM}{Large Language Model}
\acrodef{MDS}{Misbehavior Detection System}
\acrodef{RSS}{Received Signal Strength}
\acrodef{VANET}{Vehicular Ad Hoc Network}
\acrodef{WEKA}{Waikato Environment for Knowledge Analysis}
\acrodef{DS}{Dempster-Shafer}
\acrodef{EL}{Ensemble Learning}
\acrodef{IDS}{Intrusion Detection System}
\acrodef{FLDS}{False Location Detection System}
\acrodef{Ens.RF}{Ensemble Random Forest}
\acrodef{CNB}{Complement Naive Bayes}
\acrodef{DTC}{Decision Tree Classifier}
\acrodef{GBC}{Gradient Boosting Classifier}
\acrodef{MDF}{Misbehavior Detection Framework}
\acrodef{MLP}{Multi-Layer Perceptron}
\acrodef{FL}{Federated Learning}
\acrodef{RL}{Reinforcement Learning}
\acrodef{DML}{Deep Multimodal Learning}
\acrodef{IoV}{Internet of Vehicle}
\acrodef{GRU}{Gated Recurrent Unit}

\acrodef{AAA}{Authentication, Authorization and Accounting}
\acrodef{ACC}{Adaptive Cruise Control}
\acrodef{ACL}{Access Control List}
\acrodef{AD}{Automated Driving}
\acrodef{ADAM}{Adaptive Moment Estimation}
\acrodef{ADAS}{Advanced Driver Assistance Systems}
\acrodef{AKI}{Accountable Key Infrastructure}
\acrodef{API}{Application Programming Interface}
\acrodef{ASS}{Anonymity Set Size}
\acrodef{ATI}{Attack Time Impact}
\acrodef{AUC}{Area Under the Curve}
\acrodef{BSM}{Basic Safety Message}
\acrodef{BIC}{Bayesian Information Criterion}
\acrodef{BYOD}{Bring Your Own Device}
\acrodef{C2C-CC}{Car2Car Communication Consortium}
\acrodef{C2C}{Car-to-Car}
\acrodef{C2I}{Car-to-Infrastructure}
\acrodef{CA}{Certification Authority}
\acrodef{CAN}{Controller Area Network}
\acrodef{CHMM}{Continuous Hidden Markov Model}
\acrodef{CMIX}{Cryptographic Mix-Zone}
\acrodef{CN}{Common Name}
\acrodef{CVS}{Constant Vehicle Spacing}
\acrodef{CACC}{Cooperative Adaptive Cruise Control}
\acrodef{CAM}{Cooperative Awareness Message}
\acrodef{CAMP VSC3}{Crash Avoidance Metrics Partnership Vehicle Safety Consortium}
\acrodef{CAV}{Cooperative Aware Vehicle}
\acrodef{CDG}{Constant-Distance Gap}
\acrodef{CIA}{Confidentiality, Integrity and Availability}
\acrodef{CPU}{Central Processing Unit}
\acrodef{CRL}{Certificate Revocation List}
\acrodef{CTG}{Constant-Time Gap}
\acrodef{CTH}{Constant Time Headway}
\acrodef{CDN}{Content Delivery Network}
\acrodef{C-ITS}{Cooperative Intelligent Transport System}
\acrodef{COCA}{Cornell OnLine Certification Authority}
\acrodef{CSR}{Certificate Signing Request}
\acrodef{DAA}{Direct Anonymous Attestation}
\acrodef{DDoS}{Distributed Denial of Service}
\acrodef{DDH}{Decisional Diffie-Helman}
\acrodef{DENM}{Decentralized Environmental Notification Message}
\acrodef{DHMM}{Discrete Hidden Markov Model}
\acrodef{DHT}{Distributed Hash Table}
\acrodef{DL/ECIES}{Discrete Logarithm and Elliptic Curve Integrated Encryption Scheme}
\acrodef{DoS}{Denial of Service}
\acrodef{DoT}{Department of Transportation}
\acrodef{DPA}{Data Protection Agency}
\acrodef{DSRC}{Dedicated Short Range Communication}
\acrodef{DSS}{Digital Signature Standard}
\acrodef{ECU}{Electronic Control Unit}
\acrodef{EDR}{Event Data Recorder}
\acrodef{ETSI}{European Telecommunications Standards Institute}
\acrodef{ECDSA}{Elliptic Curve Digital Signature Algorithm}
\acrodef{ECC}{Elliptic Curve Cryptography}
\acrodef{EVITA}{E-safety Vehicle Intrusion protected Applications}
\acrodef{FOT}{Field Operational Test}
\acrodef{FPGA}{Field-Programmable Gate Array}
\acrodef{GPA}{Global Passive Adversary}
\acrodef{GPS}{Global Positioning System}
\acrodef{GN}{GeoNetworking}
\acrodef{GS-VLR}{Group Signatures with Verifier Local Revocation}
\acrodef{GS}{Group Signatures}
\acrodef{GM}{Group Manager}
\acrodef{GBA}{Generic Bootstrapping Architecture}
\acrodef{GNSS}{Global Navigation Satellite System}
\acrodef{GUI}{Graphic User Interface}
\acrodef{HMM}{Hidden Markov Model}
\acrodef{GMM}{Gaussian Mixture Model}
\acrodef{GMMHMM}{Gaussian Mixture Model Hidden Markov Model}
\acrodef{HSM}{Hardware Security Module}
\acrodef{HTTP}{Hypertext Transfer Protocol}
\acrodef{IEEE}{Institute of Electrical and Electronics Engineers}
\acrodef{IETF}{Internet Engineering Task Force}
\acrodef{IoT}{Internet of Things}
\acrodef{ITS}{Intelligent Transport Systems}
\acrodef{IT}{Information Technologies}
\acrodef{IVN}{In-Vehicle Network}
\acrodef{IMSI}{International Mobile Subscriber Identity}
\acrodef{IMEI}{International Mobile Station Equipment Identity}
\acrodef{IdP}{Identity Provider}
\acrodef{IDS}{Intrusion Detection System}
\acrodef{ISP}{Internet Service Provider}
\acrodef{LEA}{Law Enforcement Agency}
\acrodef{LCPP}{Lightweight Conditional Privacy Preservation}
\acrodef{LLR}{Log Likelihood Ratio}
\acrodef{LTC}{Long Term Certificate}
\acrodef{LTCA}{Long Term \acs{CA}}
\acrodef{MDS}{Misbehavior Detection Scheme}
\acrodef{H-LTCA}{Home-LTCA}
\acrodef{F-LTCA}{Foreign-LTCA}
\acrodef{LDAP}{Lightweight Directory Access Protocol}
\acrodef{LBS}{Location Based Service}
\acrodef{LSTM}{Long Short-Term Memory}
\acrodef{LTE}{Long Term Evolution}
\acrodef{LuST}{Luxembourg SUMO Traffic}
\acrodef{MA}{Misbehavior Authority}
\acrodef{MCP}{Maneuver Coordination Protocol}
\acrodef{MCM}{Maneuver Coordination Message}
\acrodef{MCS}{Maneuver Coordination Service}
\acrodef{MAC}{Media Access Control}
\acrodef{MCA}{Message \ac{CA}}
\acrodef{MEA}{Misbehavior Evaluation Authority}
\acrodef{MRM}{Minimum Risk Maneuver}
\acrodef{NN}{Neural Network}
\acrodef{NTP}{Network Time Protocol}
\acrodef{OBU}{On-Board Unit}
\acrodef{OEM}{Original Equipment Manufacturer}
\acrodef{OCSP}{Online Certificate Status Protocol}
\acrodef{PCA}{Pseudonym \acs{CA}}
\acrodef{PDP}{Policy Decision Point}
\acrodef{PEP}{Policy Enforcement Point}
\acrodef{PIR}{Private Information Retrieval}
\acrodef{PKC}{Public Key Cryptography}
\acrodef{PKCS}{Public Key Cryptosystem}
\acrodef{PKI}{Public-Key Infrastructure}
\acrodef{PRECIOSA}{Privacy Enabled Capability in Co-operative Systems and Safety Applications}
\acrodef{PRESERVE}{Preparing Secure Vehicle-to-X Communication Systems}
\acrodef{PRIME}{Platoon Restructuring for Incident Mitigation and Exclusion}
\acrodef{P2P}{peer-to-peer}
\acrodef{PS}{Participatory Sensing}
\acrodef{RA}{Resolution Authority}
\acrodef{RAM}{Random Access Memory}
\acrodef{REST}{Representational State Transfer}
\acrodef{RBAC}{Role Based Access Control}
\acrodef{RCA}{Root \acs{CA}}
\acrodef{RSU}{Roadside Unit}
\acrodef{RHyTHM}{RHyTHM}
\acrodef{SAML}{Security Assertion Markup Language}
\acrodef{SAS}{Sample Aggregation Service}
\acrodef{SCMS}{Security Credential Management System}
\acrodef{SCORE@F}{Système COopératif Routier Expérimental Français}
\acrodef{SDSI}{Simple Distributed Security Infrastructure}
\acrodef{SRAAC}{Secure Revocable Anonymous Authenticated Inter-Vehicle Communication}
\acrodef{SeVeCom}{Secure Vehicle Communication}
\acrodef{SAE}{Society of Automotive Engineers}
\acrodef{SIT}{Sichere Informationstechnologie}
\acrodef{SLC}{Short-Lived Certificate}
\acrodef{SoA}{Service-oriented-Approach}
\acrodef{SIFS}{Short Inter Frame Space}
\acrodef{SSO}{Single-Sign-On}
\acrodef{SSL}{Secure Sockets Layer}
\acrodef{SOAP}{Simple Object Access Protocol}
\acrodef{STRP}{Spatial Time Reservation Procedure}
\acrodef{SVM}{Support Vector Machine}
\acrodef{TA}{Transition Area}
\acrodef{TACK}{Temporary Anonymous Certified Key}
\acrodef{TFLite}{TensorFlow Lite}
\acrodef{TS}{Task Service}
\acrodef{TLS}{Transport Layer Security}
\acrodef{ToC}{Transition of Control}
\acrodef{TPM}{Trusted Platform Module}
\acrodef{TTP}{Trusted Third Party}
\acrodef{TVR}{Ticket Validation Repository}
\acrodef{URI}{Uniform Resource Identifier}
\acrodef{UML}{Unified Modeling Language}
\acrodef{VANET}{Vehicular Ad-hoc Network}
\acrodef{V2I}{Vehicle-to-Infrastructure}
\acrodef{V2V}{Vehicle-to-Vehicle}
\acrodef{V2X}{Vehicle-to-Everything}
\acrodef{VC}{Vehicular Communication}
\acrodef{VM}{Virtual Machine}
\acrodef{VSS}{\ac{VC} Security Subsystem}
\acrodef{WAVE}{Wireless Access in Vehicular Environments}
\acrodef{WSDL}{Web Services Discovery Language}
\acrodef{W3C}{World Wide Web Consortium}
\acrodef{V}{Vehicle}
\acrodef{VANET}{Vehicular Ad-hoc Network}
\acrodef{VLR}{Verifier-Local Revocation}
\acrodef{VPKI}{Vehicular Public-Key Infrastructure}
\acrodef{VM}{Virtual Machine}
\acrodef{WS}{Web Service}
\acrodef{WoT}{Web of Trust}
\acrodef{WSACA}{\ac{WAVE} Service Advertisement \ac{CA}}
\acrodef{XML}{Extensible Markup Language}
\acrodef{XACML}{eXtensible Access Control Markup Language}
\acrodef{ZKP}{Zero Knowledge Proof}
\acrodef{3G}{3rd Generation}
\acrodef{ROC}{Receiver Operating Characteristic}
\acrodef{TP}{True Positive}
\acrodef{FP}{False Positive}
\acrodef{TN}{True Negative}
\acrodef{FN}{False Negative}

\copyrightyear{2025}
\acmYear{2025}
\setcopyright{cc}
\setcctype{by-nc-sa}
\acmConference[WiseML 2025]{Proceedings of the 2025 ACM Workshop on Wireless Security and Machine Learning}{July 3, 2025}{Arlington, VA, USA}
\acmBooktitle{Proceedings of the 2025 ACM Workshop on Wireless Security and Machine Learning (WiseML 2025), July 3, 2025, Arlington, VA, USA}
\acmDOI{10.1145/3733965.3733966}
\acmISBN{979-8-4007-1531-0/2025/06}

\acmSubmissionID{wiseml02}
\begin{document}

\title{AttentionGuard: Transformer-based Misbehavior Detection for Secure Vehicular Platoons}

\author{Hexu Li}
\authornote{Equally contributing authors.}
\orcid{0009-0004-2430-1374}
\affiliation{%
  \institution{Networked Systems Security (NSS) Group\\KTH Royal Institute of Technology}
  \city{Stockholm}
  \country{Sweden}
}
\email{hexu@kth.se}

\author{Konstantinos Kalogiannis}
\authornotemark[1]
\orcid{0000-0002-4656-2565}
\affiliation{%
  \institution{Networked Systems Security (NSS) Group\\KTH Royal Institute of Technology}
  \city{Stockholm}
  \country{Sweden}
}
\email{konkal@kth.se}

\author{Ahmed Mohamed Hussain}
\authornotemark[1]
\orcid{0000-0003-4732-9543}
\affiliation{%
  \institution{Networked Systems Security (NSS) Group\\KTH Royal Institute of Technology}
  \city{Stockholm}
  \country{Sweden}
}
\email{ahmhus@kth.se}

\author{Panos Papadimitratos}
\orcid{0000-0002-3267-5374}
\affiliation{%
  \institution{Networked Systems Security (NSS) Group\\KTH Royal Institute of Technology}
  \city{Stockholm}
  \country{Sweden}
}
\email{papadim@kth.se}

\renewcommand{\shortauthors}{H. Li, K. Kalogiannis, A. M. Hussain, and P. Papadimitratos}

\begin{abstract}
Vehicle platooning, with vehicles traveling in close formation coordinated through Vehicle-to-Everything (V2X) communications, offers significant benefits in fuel efficiency and road utilization. However, it is vulnerable to sophisticated falsification attacks by authenticated insiders that can destabilize the formation and potentially cause catastrophic collisions. This paper addresses this challenge: misbehavior detection in vehicle platooning systems. We present \emph{AttentionGuard}, a transformer-based framework for misbehavior detection that leverages the self-attention mechanism to identify anomalous patterns in mobility data. Our proposal employs a multi-head transformer-encoder to process sequential kinematic information, enabling effective differentiation between normal mobility patterns and falsification attacks across diverse platooning scenarios, including steady-state (no-maneuver) operation, join, and exit maneuvers. Our evaluation uses an extensive simulation dataset featuring various attack vectors (constant, gradual, and combined falsifications) and operational parameters (controller types, vehicle speeds, and attacker positions). Experimental results demonstrate that \emph{AttentionGuard} achieves up to 0.95 F1-score in attack detection, with robust performance maintained during complex maneuvers. Notably, our system performs effectively with minimal latency (100ms decision intervals), making it suitable for real-time transportation safety applications. Comparative analysis reveals superior detection capabilities and establishes the transformer-encoder as a promising approach for securing  Cooperative Intelligent Transport Systems (C-ITS) against sophisticated insider threats.
\end{abstract}



\maketitle

\section{Introduction}
\label{sec:introduction}

\acp{VANET} and \ac{C-ITS} have emerged as promising technologies to enhance road safety, improve traffic efficiency, and provide a more comfortable driving experience~\cite{papadimitratos2008secure, amoozadeh2015security}. An important application in this domain is vehicle platooning, where multiple vehicles travel in close formation with reduced inter-vehicle distances, coordinated through \ac{V2X} communications~\cite{Ploeg2011CACC, Santini2017TVT}. While platooning offers substantial benefits, such as reduced fuel consumption, improved traffic throughput, and enhanced safety, it also introduces significant security vulnerabilities that must be addressed~\cite{kalogiannis2022attack}.

The vehicle platooning security challenges stem from the dependence on the continuous exchange of \acp{CAM} containing kinematic data (i.e., position, speed, acceleration) among vehicles. Cryptographic approaches provide a first line of defense, mitigating external to the platoon attackers and safeguarding against Sybil adversaries~\cite{khodaei2018Secmace}; but they remain ineffective against insider threats--authenticated vehicles that deliberately transmit falsified mobility data~\cite{Heijden2017Analyzing}. Such falsification attacks can destabilize platoon formations and potentially lead to catastrophic collisions, particularly during coordinated maneuvers, such as joining or exiting a platoon~\cite{kalogiannis2022attack}.

\acp{MDS} are an essential second line of defense, identifying anomalous patterns in \ac{V2X} communications that may indicate malicious activity~\cite{kamel2020simulation, 8885740}. Traditional \ac{MDS} approaches typically rely on pre-defined rules, thresholds, or plausibility checks, which, while computationally efficient, often fail to adapt to the complex dynamics of vehicular environments. More recent approaches leverage \ac{ML} approaches to improve detection capabilities~\cite{kamel2020simulation}; however, many challenges persist, including (i) the ability to distinguish between legitimate maneuvers and malicious behavior~\cite{kalogiannis2022attack}, (ii) computational constraints on resource-limited \acp{OBU}~\cite{9656118}, and (iii) the need for real-time misbehavior detection to enable timely mitigation~\cite{kalogiannis2024prime}.

In this paper, we propose a novel transformer-based architecture for misbehavior detection in vehicular platoons that addresses the aforementioned challenges. Our approach leverages the temporal context awareness of transformer encoders to effectively differentiate between normal mobility patterns and sophisticated falsification attacks, even during complex maneuvers. 

\textbf{Contributions.} We summarize the main contributions of this work as follows: (i) A reliable and timely \ac{MDS} capable of identifying malicious behavior in diverse mobility scenarios, including steady-state and maneuvering conditions. (ii) Evidence on the capability of transformer-based approaches to identify attacks even when the inference-phase data input differs from the training dataset. (iii) Insights on the deployment of transformer-based \acp{MDS}, in vehicles or on the edge, depending on acceptable reaction times and detection performance.

\textbf{Paper Organization.} Sec.~\ref{sec:background} reviews related work on misbehavior detection in vehicular networks, focusing on \ac{ML} approaches for platooning security. Sec.~\ref{sec:adversary-model} presents our system and adversary model, detailing platooning environments and attack scenarios. Sec.~\ref{sec:solution} describes our transformer-based framework, including the preprocessing, architecture, and implementation details. Sec.~\ref{sec:performance} evaluates our approach and analyzes practical deployment considerations for vehicle and infrastructure-based detection systems. Finally, Sec.~\ref{sec:conclusion} concludes with key findings and future research directions.
\section{Related Work}
\label{sec:background}
Vehicle platooning systems rely on reliable \ac{V2X} communications and continuous \ac{CAM} exchange. The coordinated nature of platooning operations (e.g., middle-join, exit)~\cite{kalogiannis2022attack, kalogiannis2024prime} necessitates robust misbehavior detection mechanisms to ensure operational integrity and satisfy safety-critical requirements. Traditional \ac{ML} approaches demonstrated significant efficacy in \ac{V2X} misbehavior detection~\cite{kamel2020simulation}. These include \ac{SVM}, \ac{DT}, \ac{RF}, and \ac{k-NN}, enabling systematic identification of anomalous patterns in network messages and vehicle behaviors.

Such approaches heavily depend on feature engineering and model adaptation to dynamic vehicular environments. Grover et al.~\cite{9099055} developed a comprehensive feature set encompassing geographical position, acceptance range, \ac{RSS}, speed, and packet delivery metrics for misbehavior classification in \ac{VANET}. They utilized multiple classifiers, including \ac{RF} and \ac{k-NN}, and validated the feasibility of resource-constrained misbehavior detection. Additionally, Gyawali et al.~\cite{10.1007/978-3-642-22720-2_68} integrated \ac{ML} with reputation-based systems, enhancing detection accuracy through \ac{DS}-based feedback combination. Ercan et al.~\cite{9656118} proposed a distributed \ac{IDS} utilizing ensemble learning techniques, demonstrating high performance through combining \ac{k-NN} and \ac{RF} classifiers.

\ac{DL} approaches have shown particular promise. Liu~\cite{9887617} implemented a \ac{LSTM}-based architecture for anomaly detection in the VeReMi dataset~\cite{van2018veremi}, improving on \ac{FP} rates compared to conventional detectors. Hsu et al.~\cite{10.1145/3491396.3506509} developed a hybrid architecture combining \ac{CNN} and \ac{LSTM} networks, achieving 95.35\% detection accuracy across multiple misbehavior attacks. Advanced collaborative approaches have further enhanced detection capabilities. Mangla et al.~\cite{MANGLA202352} proposed a fusion framework integrating \ac{SVM}, \ac{MLP}, and \ac{LSTM} classifiers, achieving 99.99\% accuracy in multi-class misbehavior detection. Gurjar et al.~\cite{Gurjar2025} addressed privacy concerns through \ac{FL}, enabling distributed model training while preserving data privacy. Recently, Transformer-based \ac{IDS} have been developed as a way to detect attacks in \acp{IVN}~\cite{Nam2021intrusion, Nguyen2023transformer, Cobilean2023anomaly}.

\textbf{Comparison with Existing Work.} Our \ac{MDS} incorporates a Transformer-encoder that captures the unique mobility characteristics of different maneuvering (attack and benign) scenarios. Compared to the Transformer-based solutions, our aim is not to detect attacks within \ac{IVN}; we detect adversarial behavior in platoon formations by analyzing the platooning vehicles' sensor readings and mobility patterns.
\section{System and Adversary Model}
\label{sec:adversary-model}

We consider a set of \ac{V2X}-enabled vehicles forming a platoon on a highway and traveling in unison. During the journey, other \ac{V2X}-enabled vehicles can request to join the formation to gain platooning benefits. The platoon leader can accept such a request and designate a join position based on the requesting car's capabilities and target destination. Similarly, platooning vehicles can decide to exit their formation during the trip, e.g., when approaching their destination, by initiating an exit request and performing the maneuver when available.

It is exactly at these moments that the attacker can choose to strike, to maximize its attack potential effect (harm). More specifically, we study a threat model applicable to \ac{VC} systems~\cite{papadimitratos2008secure}, and more specifically to platooning~\cite{amoozadeh2015security, kalogiannis2022attack}. An attacker possesses valid cryptographic credentials and can join platoons to perform its attacks: the attacker cannot impersonate another vehicle, but it can alter the kinematic properties (i.e., position, speed, acceleration) of the \acp{CAM} it transmits. Based on the dataset in~\cite{kalogiannis2022attack}, the attacks can introduce a constant or a gradual offset to each of the kinematic values; or alter them intelligently, and in unison, in a physics-consistent way, termed ``combined'' attacks. These nine types of attacks are executed against platoons in different mobility scenarios: a join, an exit, or a steady-state scenario.
\section{Proposed Framework}
\label{sec:solution}
We propose an \ac{MDS} framework consisting of three phases, namely: (i) data pre-processing, (ii) model training, and (iii) model evaluation.

\textbf{Data Pre-Processing.} The available simulated data consists of platoons of up to seven vehicles traveling on a straight highway for 118.9 seconds, recording seven mobility properties (\textit{distance}, \textit{relativeSpeed}, \textit{acceleration}, \textit{controllerAcceleration}, \textit{speed}, \textit{posx}, and \textit{posy}). To account for varying vehicle insertion times (appearing in the simulation), we utilize a \textit{mask} array indicating valid features for each car. Finally, we classify data as benign (0), or attack (1) if they deviate from the ground truth; for scenarios under the same seed, the data deviate only when an attack affects the vehicle controller. This results in an array of shape (7, 1189, 7) for the features and an array of shape (7, 1189) for the labels and the mask. Table~\ref{tab:seedStats} summarizes the total number of traces and the ratio between the benign and attack traces processed.

\begin{table}[h!]
\centering
\caption{Outcome of preprocessing for each seed.}
\label{tab:seedStats}
\resizebox{\columnwidth}{!}{%
\begin{tabular}{l c c c c c c}
\toprule
\textbf{Seed} & 
\textbf{Total Traces} & 
\textbf{Valid Labels} & 
\textbf{Ratio of 1} & 
\textbf{Ratio of 0} & 
\textbf{0/1 Ratio} \\
\midrule
0-02-1   & 1454 & 8307624 & 0.2111 & 0.7889 & 3.7369 \\
2016-02-1& 1620 & 9568553 & 0.2863 & 0.7137 & 2.4926 \\
2083-02-1& 1449 & 8286245 & 0.2217 & 0.7783 & 3.5101 \\
2084-02-1& 1458 & 8376577 & 0.2209 & 0.7791 & 3.5278 \\
2085-02-1& 1426 & 8119617 & 0.2199 & 0.7801 & 3.5469 \\
2086-02-1& 1434 & 8168407 & 0.2174 & 0.7826 & 3.6007 \\
\bottomrule
\end{tabular}%
}
\end{table}

To ensure all features contribute equally during the \textit{model} \emph{training} phase, we normalize values using a global mean $\mu$ and variance $\sigma^2$. Specifically,  let \(\mathbf{x} \in \mathbb{R}^{7} \) be the raw feature vector at a given time step.  We first compute:

\begin{equation}
\mu = \frac{1}{N} \sum_{n=1}^{N} \mathbf{x}_n, 
\quad
\sigma^2 = \frac{1}{N} \sum_{n=1}^{N} (\mathbf{x}_n - \mu)^2,
\end{equation}

where \(N\) is the total number of feature vectors across all files. Then, each feature is normalized to:

\begin{equation}
\hat{\mathbf{x}} = \frac{\mathbf{x}-\mu}{\sigma + \epsilon},
\label{eq:zscore}
\end{equation}

with \(\epsilon \approx 10^{-8}\) to avoid division by zero. 
The z-score transformation ensures that each dimension has a near-zero mean and unit variance, reducing scale discrepancies among different scenarios. After normalization, we split the data into equal-sized sliding windows (10 data points) and introduce padding where applicable.

\textbf{Considered Model.} We utilize the Transformer-encoder architecture because it can generate output data by capturing the context of the input data~\cite{vaswani2017attention}. Unlike the sequential processing of data performed by \acp{RNN}, Transformers can at once process the entire data sequence, allowing for extracting relationships and context within the sequence. The overall Transformer architecture (introduced in~\cite{vaswani2017attention}) includes two primary components: the \textit{encoder} and the \textit{decoder}. The encoder is tasked with processing the input data sequence, transforming it into a continuous representation that encapsulates contextual information. This is achieved through multiple layers, each incorporating a multi-head self-attention mechanism and a position-wise fully connected feed-forward network. The self-attention mechanism enables the encoder to assign varying degrees of importance to different parts of the input sequence, thereby generating a comprehensive representation of the data. 

\begin{figure}[!t]
    \centering
    \includegraphics[width=\columnwidth]{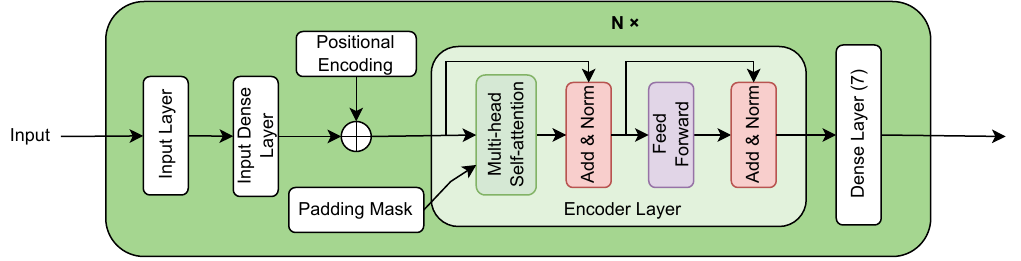}
    \caption{Structure and details of the implemented transformer-encoder.}
    \label{fig:transformer}
\end{figure}

The complete architecture of our transformer block is presented in Fig.~\ref{fig:transformer}. For this domain-specific problem, our approach consists of two model training approaches: (i) a \textit{general} platooning model utilizing all vehicle data as input, (ii) a \textit{vehicle-specific} model trained only on its own mobility data. The aim is to provide insights on the feasibility of deploying general or vehicle-specific models to ensure the vehicular formation security.

Considering the binary nature of the problem (benign vs attack characterization), we utilize the masked binary-cross entropy loss function, presented in Equation~\ref{eq:maskedBCE}.
Let $y_i \in \{0,1\}$ be the ground-truth label for step $i$, and $\mathrm{mask}_i \in \{0,1\}$  indicate whether that step is valid (i.e.\ not padded). The model outputs a logit  $\hat{z}_i \in \mathbb{R}$. We define a \textit{positive weight} $\alpha > 1$ to handle the class imbalance inherent in our data. Thus, for each valid step ($\mathrm{mask}_i=1$), the BCE loss is:

\begin{equation}
\mathcal{L}_\mathrm{BCE} = 
-\sum_{i} \mathrm{mask}_i \,\Bigl[\, 
(1-y_i)\,\log\bigl(\sigma(\hat{z}_i)\bigr) \;+\;
y_i\,\alpha\,\log\bigl(\sigma(\hat{z}_i)\bigr)\Bigr]
\label{eq:maskedBCE}
\end{equation}

where $\sigma(\cdot)$ is the sigmoid function.  All steps with $\mathrm{mask}_i=0$ do not contribute to the loss.  Finally, we normalize by the total number of valid steps to get an average loss.

\textbf{Model Evaluation.} To drive the discussion towards the location deployment of \acp{MDS}, in-vehicle, on the edge, or at a \ac{RSU}, we run the inference phase of the general platooning model with two different inputs: either locally, with just the vehicle own mobility data, or after collecting the entire platoon mobility information. Additionally, we evaluate each vehicle's model performance against the general platooning model. 

Our evaluation of attack detection employs several metrics, \emph{recall}, \emph{precision}, \emph{$F_1$ score}, and \emph{accuracy}; giving us, respectively, the fraction of true positive steps among all actual positives, the fraction of correctly identified positives among all predicted positives, the harmonic mean of the previous two metrics (useful when dealing with unbalanced classes), and, finally, the indicator of correct predictions. Furthermore, we use the \ac{ROC}, the proportion of correctly detected messages (\ac{TP}) over incorrectly identified messages (\ac{FP}), to measure the ability of our identifier to correctly classify the input data. Additionally, we provide the \ac{AUC} to facilitate the classifier comparison.
\section{Performance Evaluation}
\label{sec:performance}

\subsection{Training Setup}
\label{subsec:sampleSize}
The transformer-encoder is implemented using TensorFlow~\cite{tf} and Keras~\cite{keras}; for training, we used a batch size of 128, chosen empirically, and a learning rate of $5\epsilon^{-5}$ for the \ac{ADAM} optimizer. We set the positive weights for all models depending on the ratio of benign and attack labels, as described in Table~\ref{tab:seedStats}, e.g., to 3.3 for the general platooning model. For positional encoding, we utilize the $\sin(\cdot)$ and $\cos(\cdot)$ functions~\cite{vaswani2017attention}. As part of the training and inference process, we define the window size to be 10, i.e., we gather 1s worth of data, and consider a variable step size; from 1 to 10, signifying the required data before making the next prediction (from 100ms to 1s, respectively). Table~\ref{tab:sim-parameters} outlines all the parameters used in the model training setup.

\begin{table}[!h]
    \centering
    \caption{Training parameters.}
    \label{tab:sim-parameters}
    \resizebox{\columnwidth}{!}{
        \begin{tabular}{c c c c}
            \toprule
            \multicolumn{2}{c}{\textbf{Transformer-encoder}} & \multicolumn{2}{c}{\textbf{Platoons}}  \\
            \cmidrule(lr){1-2} \cmidrule(lr){3-4}
            \textbf{Parameters} & \textbf{Value} & \textbf{Parameters} & \textbf{Value}\\
            \midrule
            Batch Size & 128 & Cars & 6 or 7 (Join) \\
            Data Split Ratio  & 80/20 & Spacing Controller & PATH, Flatbed \\
            Learning Rate   & $5\epsilon^{-5}$ & Headway Controller & Ploeg, Consensus \\
            Window size   & 10 & Leader Speed & 50, 80, 100, 150 kmph \\
            Step size   & [1, 5, 10] & Spacing & 5m, 5m, 0.5s, 0.8s \\
            Positive Weight/Positional Encoding & Dataset/Model dependent & Sim. Duration & 120s \\
            Loss Function & Binary Crossentropy & Attacker Position & [0, 2]\\
            \bottomrule
        \end{tabular}
    }
\end{table}

On the other hand, the input dataset describes platoons of 6 vehicles, increasing to 7 when a join needs to be performed, that travel on the highway with different speeds. The platoons themselves utilize different controllers, either \ac{CVS}, where the vehicles try to keep a constant spacing distance; or \ac{CTH}, where the vehicles' intra-platoon distances are based on speed and the time it would take to reach the bumper of the car ahead. The different controllers and speeds contribute to the diverse mobility characteristics of the vehicles. Finally, we consider attackers positions either at the front, as the platoon leader, or at position three (Vehicle 2), as a regular platoon follower; the latter corresponds to the position just in front of the join or exit positions~\cite{kalogiannis2022attack}. The analysis of our data was performed on an Ubuntu machine, using 128 GB of \ac{RAM}, an AMD Ryzen Threadripper PRO 5965WX with 24 physical cores and 48 logical cores, and an NVIDIA GeForce RTX 4090 with 24 GB of DDR5 memory. 

\begin{figure}[!h]
    \centering
    \includegraphics[width=\linewidth]{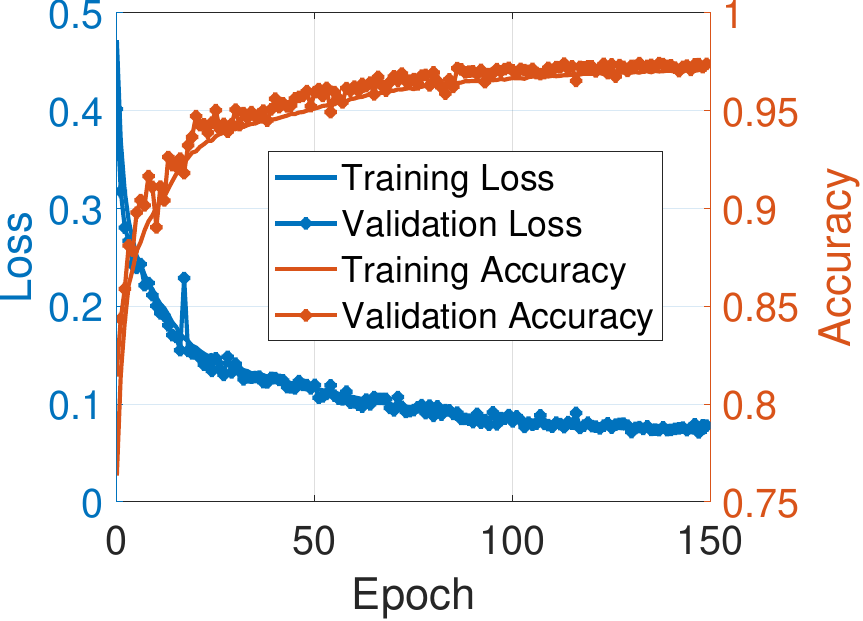}
    \caption{Model training/validation Loss/Accuracy as a function of number of epochs.}
    \label{fig:loss-accuracy}
\end{figure}

\subsection{Evaluation Results}
\label{sec:results}
Fig.~\ref{fig:loss-accuracy} illustrates the loss and accuracy values per epoch for a maximum of 150 epochs during the training phase of the general platooning model. We observe that both values steadily converge, indicating that the model can learn from our input. During the training phase, we reach an accuracy of 0.96 on the validation and training datasets. We omit the training plots for the individual car models.

\begin{figure}[!htbp]
\centering
  \begin{subfigure}[b]{0.33\textwidth}
   \includegraphics[width=\linewidth]{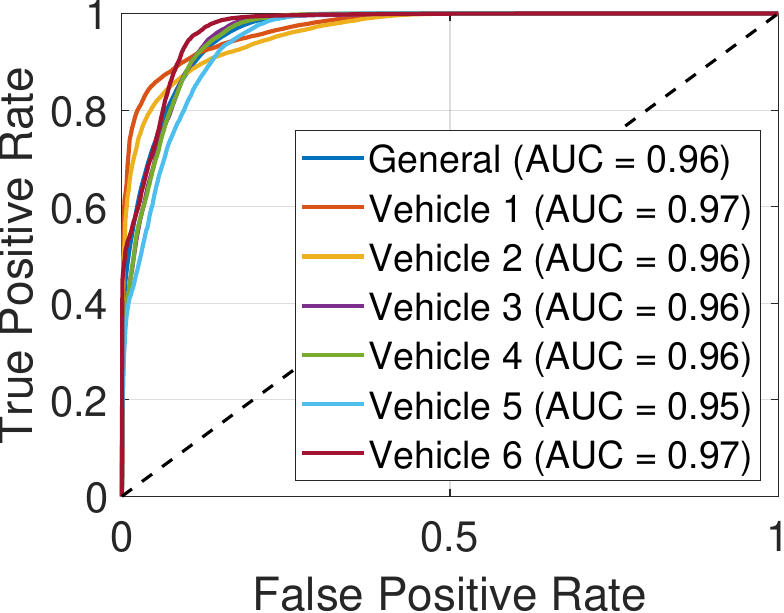}
    \caption{}
    \label{fig:roc_a}
  \end{subfigure}
  \hspace{1em}
   \begin{subfigure}[b]{0.33\textwidth}
    \includegraphics[width=\linewidth]{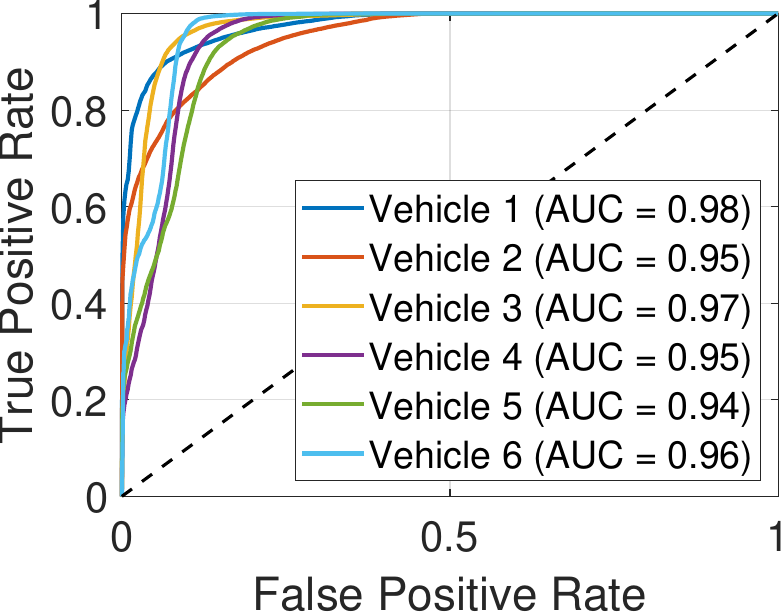}
    \caption{}
    \label{fig:roc_b}
  \end{subfigure}
    \caption{\ac{ROC} curves for step = 5: (a) Global (General) Model, (b) Vehicle Model.}
    \label{fig:roc}
\end{figure}

\begin{figure*}[!htbp]
\centering
  \begin{subfigure}[b]{0.22\textwidth}
    \includegraphics[width=\linewidth]{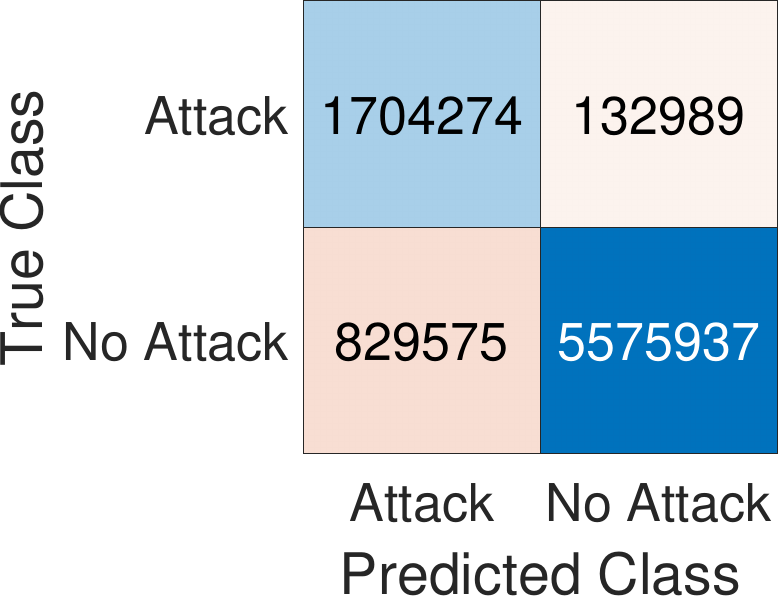}
    \caption{}
    \label{fig:cm_a}
  \end{subfigure}
  \hspace{1em}
  \begin{subfigure}[b]{0.22\textwidth}
    \includegraphics[width=\linewidth]{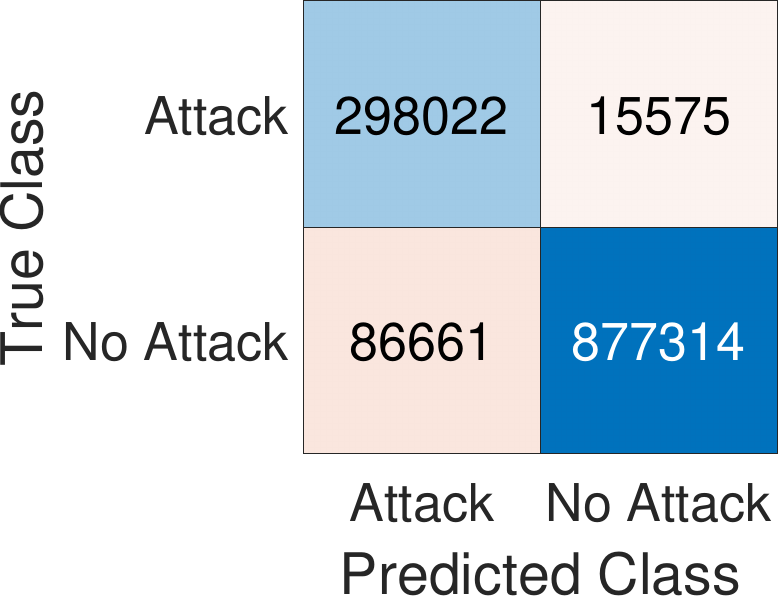}
    \caption{}
    \label{fig:cm_b}
  \end{subfigure}
  \hspace{1em}
   \begin{subfigure}[b]{0.22\textwidth}
    \includegraphics[width=\linewidth]{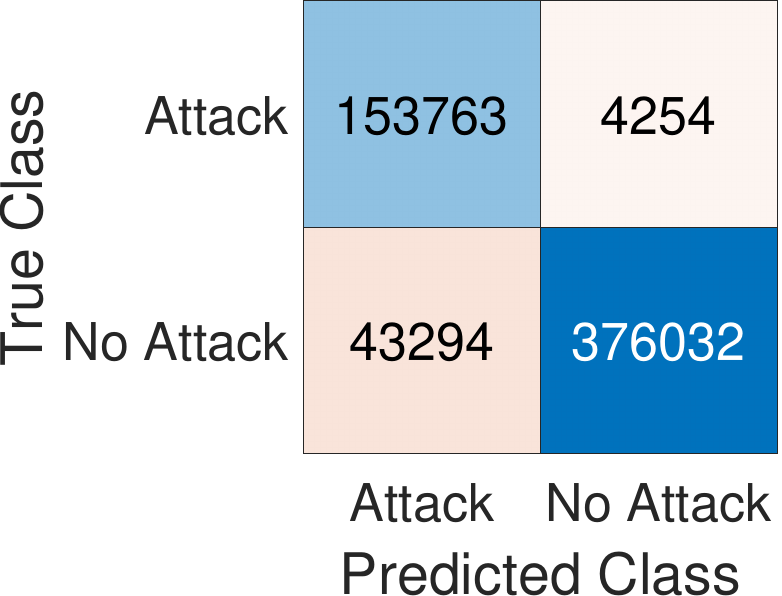}
    \caption{}
    \label{fig:cm_c}
  \end{subfigure}
    \caption{Confusion matrix for step = 5: (a) Global Model, (b) Vehicle 3 Model, and (c) Vehicle 6 Model.}
    \label{fig:cm}
\end{figure*}

\definecolor{heat-lowest}{HTML}{E74C3C}    
\definecolor{heat-low}{HTML}{F5B041}       
\definecolor{heat-medium-low}{HTML}{F7DC6F}
\definecolor{heat-medium}{HTML}{F9E79F}    
\definecolor{heat-medium-high}{HTML}{ABEBC6}
\definecolor{heat-high}{HTML}{82E0AA}      
\definecolor{heat-highest}{HTML}{2ECC71}   

\newcommand{\heatmapcell}[1]{%
  \ifdim#1pt<0.83pt
    \cellcolor{heat-lowest} #1
  \else\ifdim#1pt<0.86pt
    \cellcolor{heat-low} #1
  \else\ifdim#1pt<0.89pt
    \cellcolor{heat-medium-low} #1
  \else\ifdim#1pt<0.92pt
    \cellcolor{heat-medium} #1
  \else\ifdim#1pt<0.95pt
    \cellcolor{heat-medium-high} #1
  \else\ifdim#1pt<0.98pt
    \cellcolor{heat-high} #1
  \else
    \cellcolor{heat-highest} #1
  \fi\fi\fi\fi\fi\fi
}

\begin{table*}[!h]
\centering
\caption{Performance comparison for general platooning and individual vehicle models.}
\label{tab:car-models}
\resizebox{0.85\textwidth}{!}{%
    \begin{tabular}{c c c c c | c c c c | c c c c}
        \toprule
        \textbf{Input} & \multicolumn{4}{c}{\textbf{Step = 1}} & 
        \multicolumn{4}{c}{\textbf{Step = 5}} & \multicolumn{4}{c}{\textbf{Step = 10}}\\
        \cmidrule(lr){2-5} \cmidrule(lr){6-9} \cmidrule(lr){10-13}
        & \textbf{Accuracy} & \textbf{Precision} & \textbf{Recall} & \textbf{F1-score}
        & \textbf{Accuracy} & \textbf{Precision} & \textbf{Recall} & \textbf{F1-score}
        & \textbf{Accuracy} & \textbf{Precision} & \textbf{Recall} & \textbf{F1-score} \\
        \hline
    General & \heatmapcell{0.88} & \heatmapcell{0.91} & \heatmapcell{0.88} & \heatmapcell{0.89} & \heatmapcell{0.88} & \heatmapcell{0.91} & \heatmapcell{0.88} & \heatmapcell{0.89} & \heatmapcell{0.91} & \heatmapcell{0.93} & \heatmapcell{0.91} & \heatmapcell{0.91} \\
    Vehicle 1 & \heatmapcell{0.90} & \heatmapcell{0.92} & \heatmapcell{0.90} & \heatmapcell{0.90} & \heatmapcell{0.90} & \heatmapcell{0.93} & \heatmapcell{0.90} & \heatmapcell{0.91} & \heatmapcell{0.95} & \heatmapcell{0.95} & \heatmapcell{0.95} & \heatmapcell{0.95} \\
    Vehicle 2 & \heatmapcell{0.84} & \heatmapcell{0.89} & \heatmapcell{0.84} & \heatmapcell{0.85} & \heatmapcell{0.85} & \heatmapcell{0.90} & \heatmapcell{0.85} & \heatmapcell{0.86} & \heatmapcell{0.91} & \heatmapcell{0.93} & \heatmapcell{0.91} & \heatmapcell{0.91} \\
    Vehicle 3 & \heatmapcell{0.92} & \heatmapcell{0.93} & \heatmapcell{0.92} & \heatmapcell{0.92} & \heatmapcell{0.92} & \heatmapcell{0.93} & \heatmapcell{0.92} & \heatmapcell{0.92} & \heatmapcell{0.94} & \heatmapcell{0.95} & \heatmapcell{0.94} & \heatmapcell{0.94} \\
    Vehicle 4 & \heatmapcell{0.89} & \heatmapcell{0.90} & \heatmapcell{0.89} & \heatmapcell{0.89} & \heatmapcell{0.90} & \heatmapcell{0.91} & \heatmapcell{0.90} & \heatmapcell{0.90} & \heatmapcell{0.93} & \heatmapcell{0.93} & \heatmapcell{0.93} & \heatmapcell{0.93} \\
    Vehicle 5 & \heatmapcell{0.87} & \heatmapcell{0.89} & \heatmapcell{0.87} & \heatmapcell{0.87} & \heatmapcell{0.88} & \heatmapcell{0.89} & \heatmapcell{0.88} & \heatmapcell{0.88} & \heatmapcell{0.90} & \heatmapcell{0.91} & \heatmapcell{0.90} & \heatmapcell{0.90} \\
    Vehicle 6 & \heatmapcell{0.92} & \heatmapcell{0.93} & \heatmapcell{0.92} & \heatmapcell{0.92} & \heatmapcell{0.92} & \heatmapcell{0.93} & \heatmapcell{0.92} & \heatmapcell{0.92} & \heatmapcell{0.92} & \heatmapcell{0.93} & \heatmapcell{0.92} & \heatmapcell{0.92} \\
    \bottomrule
    \end{tabular}
}
\end{table*}

We consider model input, with a step equal to 5, a balance between timely predictions and detection performance (shown in Tables~\ref{tab:car-models} and~\ref{tab:general-models}). Fig.~\ref{fig:roc} presents the \ac{ROC} curves comparing the detection performance of \textit{AttentionGuard}. In Fig.~\ref{fig:roc_a}, the general platoon-level model (with the platoon input) achieves an \ac{AUC} of 0.96, demonstrating robust classification capabilities. Individual vehicle inputs show similar but lower performance, with Vehicles 3 and 6 exhibiting better detection rates at lower \ac{FP} rates. In comparison, vehicles 1 and 2 show steeper initial positive rates, but with a lower convergence as the input increases. Vehicle 0 (platoon leader) is not shown, as the attacks have no effect, and the classifier results in $>$0.99 across the evaluation metrics. In Fig.~\ref{fig:roc_b}, we showcase the performance of the individual car models: the majority perform better, but notably, vehicles 2 and 5 show higher \ac{FP} rates. These vehicles are the furthest away from both attacking vehicles (0 and 2, respectively), making distinguishing between misbehavior and benign movement harder. Across all models, these consistently high \ac{AUC} values validate the transformer architecture capability to capture attack patterns, making it an effective approach for misbehavior detection in vehicle platooning scenarios. 

Fig.~\ref{fig:cm} presents the confusion matrices for three different models utilizing a step 5 input, i.e., 500ms worth of new data. The platoon-level model, in Fig.~\ref{fig:cm_a} has 1,704,274 \acp{TP} and 5,575,937 \acp{TN}, indicating effective identification of non-attack instances across the entire formation. However, the model does not avoid misclassifications, with 829,575 \acp{FP} and 132,989 \acp{FN}; affecting the precision of the attack detection. This translates to an accuracy of 0.88, with a weighted precision of 0.91, recall of 0.88, resulting in a F1-score of 0.89 as detailed in Table~\ref{tab:car-models}. 

In Fig.~\ref{fig:cm_b} and Fig.~\ref{fig:cm_c}, we show the confusion matrix for vehicles 3 and 6 respectively (i.e., in each case a vehicle-specific model). These correspond to the vehicle traveling behind the follower-attacker and the vehicle entering the formation just behind it, respectively, a prime target to cause harm to both the platoon downstream and vehicles on the next lane. For vehicle 3, the confusion matrix contains a smaller number of benign and attack samples; however, the classifier improves substantially, with an increase of 0.2-0.4 across all metrics, even during maneuvering. For vehicle 6, the model demonstrates similar improvements (from the general model), illustrating the effectiveness of our approach, even for the joining vehicle. 

\begin{table*}[!t]
\centering
\caption{Performance comparison for the general platooning model with vehicle-specific input.}
\label{tab:general-models}
\resizebox{0.85\textwidth}{!}{%
    \begin{tabular}{c c c c c | c c c c | c c c c}
        \toprule
        \textbf{Input} & \multicolumn{4}{c}{\textbf{Step = 1}} & 
        \multicolumn{4}{c}{\textbf{Step = 5}} & \multicolumn{4}{c}{\textbf{Step = 10}}\\
        \cmidrule(lr){2-5} \cmidrule(lr){6-9} \cmidrule(lr){10-13}
        & \textbf{Accuracy} & \textbf{Precision} & \textbf{Recall} & \textbf{F1-score}
        & \textbf{Accuracy} & \textbf{Precision} & \textbf{Recall} & \textbf{F1-score}
        & \textbf{Accuracy} & \textbf{Precision} & \textbf{Recall} & \textbf{F1-score} \\
        \hline
        General & \heatmapcell{0.88} & \heatmapcell{0.91} & \heatmapcell{0.88} & \heatmapcell{0.89} & \heatmapcell{0.88} & \heatmapcell{0.91} & \heatmapcell{0.88} & \heatmapcell{0.89} & \heatmapcell{0.91} & \heatmapcell{0.93} & \heatmapcell{0.91} & \heatmapcell{0.91} \\
        Vehicle 1 & \heatmapcell{0.83} & \heatmapcell{0.90} & \heatmapcell{0.83} & \heatmapcell{0.85} & \heatmapcell{0.84} & \heatmapcell{0.91} & \heatmapcell{0.84} & \heatmapcell{0.85} & \heatmapcell{0.89} & \heatmapcell{0.93} & \heatmapcell{0.89} & \heatmapcell{0.90} \\
        Vehicle 2 & \heatmapcell{0.80} & \heatmapcell{0.89} & \heatmapcell{0.80} & \heatmapcell{0.82} & \heatmapcell{0.80} & \heatmapcell{0.90} & \heatmapcell{0.80} & \heatmapcell{0.82} & \heatmapcell{0.85} & \heatmapcell{0.91} & \heatmapcell{0.85} & \heatmapcell{0.86} \\
        Vehicle 3 & \heatmapcell{0.89} & \heatmapcell{0.91} & \heatmapcell{0.89} & \heatmapcell{0.90} & \heatmapcell{0.89} & \heatmapcell{0.91} & \heatmapcell{0.89} & \heatmapcell{0.90} & \heatmapcell{0.92} & \heatmapcell{0.93} & \heatmapcell{0.92} & \heatmapcell{0.92} \\
        Vehicle 4 & \heatmapcell{0.89} & \heatmapcell{0.90} & \heatmapcell{0.89} & \heatmapcell{0.89} & \heatmapcell{0.90} & \heatmapcell{0.90} & \heatmapcell{0.90} & \heatmapcell{0.90} & \heatmapcell{0.92} & \heatmapcell{0.92} & \heatmapcell{0.92} & \heatmapcell{0.92} \\
        Vehicle 5 & \heatmapcell{0.88} & \heatmapcell{0.89} & \heatmapcell{0.88} & \heatmapcell{0.88} & \heatmapcell{0.88} & \heatmapcell{0.89} & \heatmapcell{0.88} & \heatmapcell{0.88} & \heatmapcell{0.90} & \heatmapcell{0.91} & \heatmapcell{0.90} & \heatmapcell{0.90} \\
        Vehicle 6 & \heatmapcell{0.90} & \heatmapcell{0.92} & \heatmapcell{0.90} & \heatmapcell{0.91} & \heatmapcell{0.91} & \heatmapcell{0.92} & \heatmapcell{0.91} & \heatmapcell{0.91} & \heatmapcell{0.91} & \heatmapcell{0.92} & \heatmapcell{0.91} & \heatmapcell{0.92} \\
    \bottomrule
    \end{tabular}
}
\end{table*}

\subsection{\ac{MDS} Deployment}
\label{subsec:deployment}
An essential part of any vehicular \ac{MDS}, particularly in platooning where vehicles travel closely together, is to swiftly and reliably determine any abnormal mobility. In Tables~\ref{tab:car-models} and~\ref{tab:general-models}, we observe that with smaller step sizes, the \ac{MDS} trades off worse performance across all evaluation metrics, for faster reaction to any detected misbehavior. A step size of 10 can potentially be catastrophic for the platoon, making a faster but less precise decision preferable when coupled with an adequate mitigation response~\cite{kalogiannis2024prime}. 

In Table~\ref{tab:car-models}, we evaluate the performance of the different models during inference, using weighted averages. The general platooning model confirms the result in Fig.~\ref{fig:roc_b}), by performing slightly better than the individual models of vehicles 2 and 5. Nonetheless, the rest of the individual models show an increase for all metrics; specifically, the F1-score improves by 0.1-0.3 for the step equal to 1; 0.2-0.3 for the step equal to 5; and 0.1-0.4, reaching a total of 0.95 for vehicle 1. This can be anticipated as individual models were trained specifically for each vehicle. Comparing the individual vehicle performance, vehicles closest to the attackers achieve the best detection rates, even compared to the general model. 

For the general platooning model (Table~\ref{tab:general-models}), individual vehicle inputs result in varying performance, depending on the position of the vehicles in the formation. Vehicles 1 and 2 are unaffected by all follower attacks (performed by Vehicle 2 itself), as they are upstream and can only detect malicious behavior of a leader attacker (Vehicle 0). Resulting in degraded \ac{MDS} performance, highlighted in red (with a downward trend for smaller step size), as the model is trained on the full dataset and not just their behavior. Both vehicles show a high number of \acp{FP}, affecting the attack-class precision, and \acp{FN}, affecting the recall. Notably, a vehicle that decides to enter the platoon formation (Vehicle 6 in our simulations) performs well according to all the metrics (lowest being 0.90), for a step equal to 1. This means that the joiner can immediately distinguish misbehavior, avoid joining the formation and getting affected, thus avoiding causing hazardous conditions to the platoon itself, or other non-platooning vehicles on the road.

\textbf{Discussion.} Examining both model training approaches (general platooning and vehicle-specific), the individual models generally perform better (except for vehicles 2 and 5). However, this requires vehicles having different models depending on their position in a platoon, effectively limiting them to platoons of size equal to their number of models. With a general platooning model, individual inference inputs provide comparable results to the entire platoon input, except for vehicles 1 and 2. This implies that training and inference on all the platoons' data can effectively safeguard a platoon by discerning platoon misbehavior. This allows for a more flexible deployment of \acp{MDS}, e.g., on an \ac{RSU} or the platoon leader, enabling it to detect misbehavior within the platoon. Further, given higher step sizes, when timely decisions may not be critical, such an \ac{MDS} could act as a forensic tool for post-attack analysis. 

However, deploying \emph{AttentionGuard} on a vehicle or an edge device (representing an \ac{RSU}) presents several challenges. Notably, the limited computational power and memory constraints of such platforms. Towards addressing these challenges, we propose using optimization methods provided by the TensorFlow library. Namely, model conversion to \ac{TFLite} and quantization. Such optimizations can significantly reduce the model size and improve inference time, enabling deployment without accuracy degradation~\cite{hussain2024edge}.
\section{Conclusion}
\label{sec:conclusion}
We presented \emph{AttentionGuard}, a transformer-based \ac{MDS} framework to safeguard vehicle platoon formations, operating under different controllers, speeds, and maneuvering states. Our evaluation shows that our approach can provide high detection rates and fast reaction, allowing vehicles to detect attacks even after 100ms. Further, we presented results that support the training of models on entire platoon data, while still guaranteeing the detection performance of individual cars when deployed locally. This enables diverse configurations depending on the preferred outcome. As part of our future work, we will expand our detection scheme to cover a plethora of network attacks in platooning and apply model optimizations for deployment on resource-constrained devices while analyzing their overall performance.

\begin{acks}
This work is supported in parts by the Swedish Research Council (VR) and the Knut and Alice Wallenberg (KAW) Foundation.
\end{acks}

\balance
\bibliographystyle{ACM-Reference-Format}
\bibliography{main}

\end{document}